\documentclass[%
reprint,
superscriptaddress,
nofootinbib,
amsmath,amssymb,
aps,
prb,
floatfix,
longbibliography
]{revtex4-2}

\usepackage{graphicx}
\usepackage{dcolumn}
\usepackage{bm}
\usepackage[colorlinks=true, linkcolor=blue,urlcolor=blue,citecolor=blue]{hyperref}
\usepackage[mathlines]{lineno}

\usepackage[ignoreunlbld,norefs,nocites]{refcheck}

\usepackage{soul}

\usepackage{ulem}

\begin{document}

\preprint{APS/123-QED}

\title{Wire width and density dependence of the crossover in the peak of the static structure factor from $2k_\text{F}$ $\rightarrow$
  $4k_\text{F}$ in one-dimensional paramagnetic electron gases}
\author{Ankush Girdhar}
\affiliation{%
Department of Physics, Dr.\ B.\ R.\ Ambedkar National Institute of Technology, Jalandhar, Punjab-144027, India}%

\author{Vinod Ashokan}
\email{ashokanv@nitj.ac.in}
\affiliation{Department of Physics, Dr.\ B.\ R.\ Ambedkar National Institute of Technology, Jalandhar, Punjab-144027, India}
\author{Rajesh O. Sharma}
\affiliation{Department of Physics, Indian Institute of Science, Bengaluru-560012, India.}%

\author{N.\ D.\ Drummond}
\affiliation{Department of Physics, Lancaster University, Lancaster LA1 4YB, United Kingdom}%

\author{K.\ N.\ Pathak}%
\affiliation{Centre for Advanced Study in Physics, Panjab
 University, Chandigarh-160014, India}%

\date{\today}

\begin{abstract}
We use the variational quantum Monte Carlo (VMC) method to study the wire width ($b$) and electron density ($r_\text{s}$) dependences of the ground-state properties of quasi-one-dimensional paramagnetic electron fluids.
The onset of a quasi-Wigner crystal phase is known to depend on electron density, and the crossover occurs in the low density regime. We study the effect of wire width on the crossover of the dominant peak in the static structure factor from $k=2k_\text{F}$ to $k=4k_\text{F}$. It is found that for a fixed electron density, in the charge structure factor the crossover from the dominant peak occurring at $2k_\text{F}$ to $4k_\text{F}$ occurs as the wire width decreases. Our study suggests that the crossover is due to interplay of both $r_\text{s}$ and $b<r_\text{s}$. The finite wire width correlation effect is reflected in the peak height of the charge and spin structure factors. We fit the dominant peaks of the charge and spin structure factors assuming fit functions based on our finite wire width theory and clues from bosonization, resulting in a good fit of the VMC data. The pronounced peaks in the charge and spin structure factors at $4 k_\text{F}$ and $2 k_\text{F}$, respectively, indicate the complete decoupling of the charge and spin degrees of freedom. Furthermore, the wire width dependence of the electron correlation energy and the Tomonaga-Luttinger parameter $K_{\rho}$ is found to be significant. 

\end{abstract}

\maketitle


\section{Introduction}

Many-body correlation effects in one-dimensional (1D) electron systems give rise to fascinating properties with diverse technological applications \cite{Giuliani_2005, Giamarchi04}. Advances in fabrication techniques have made the realization of extremely thin wires feasible, although only a few aspect of theoretical predictions have yet been experimentally tested. This is due to the experimental difficulties of realizing ideal and controllable 1D systems. There are several features of 1D systems, such as spin-charge separation \cite{Auslaender05}, power-law behavior of correlation functions and other physical quantities such as conductance \cite{Bockrath99}, charge fractionalization \cite{Steinberg08}, Wigner crystallization \cite{Deshpande08, Pecker13, Shapir19, Ziani20}, etc., which make them unique and interesting. The physics of 1D electron systems cannot be explained using Landau's Fermi liquid theory due to the fact that single-particle excitation energies and their inverse lifetimes are of the same order of magnitude. 

The exactly solvable model by Tomonaga and Luttinger \cite{Tomonaga_1950, Luttinger_1963, Haldane81} describes the low-energy spectrum of a 1D homogeneous electron gas (HEG) assuming short-range interactions and a linear dispersion relation. The assumption of short-range interactions is quite successfully used in studies of the properties of various 1D structures such as quasi-one-dimensional conductors where screening between adjacent chains leads to effective short-range interactions within each chain \cite{Schulz83}. However, the electrons in systems which do not have effective screening, such as isolated metallic carbon nanotubes \cite{Shapir19}, interact via the true long-range Coulomb potential $[V(x)=e^2/|x|]$. 
For the purpose of theoretical studies of 1D electron systems, there are several choices of confinement model available such as hard wall \cite{Gold90, Hu93}, harmonic \cite{Friesen80, Hu90}, and Coulombic \cite{Sun93}. In all these cases, the effective electron-electron interaction is known to have a $1/|x|$ long range tail in the lowest subband of the transverse motion. In this study, we consider electrons interacting through an effective long-range Coulomb interaction using a harmonic confinement model. 

1D interacting electron systems have been studied using several techniques such as density functional theory (DFT) \cite{Sun93}, exact diagonalization \cite{Poilblanc97}, the random phase approximation (RPA) \cite{Renu14, Vinod18, Morawetz18, Vinod20,Friesen80, Camels97}, density matrix renormalization group (DMRG) \cite{Fano99, Li19}, quantum Monte Carlo (QMC) \cite{Malatesta00, Casula05, Lee11, Vinod18c, Ankush22}, lattice regularized diffusion Monte Carlo (LRDMC) \cite{ Casula06, Shulenburger08} and bosonization \cite{Schulz93}. The latter was used by Schulz in an interesting paper \cite{Schulz93} studying a 1D electron gas with long range interactions assuming a linearized kinetic energy dispersion. The major finding of the paper was an extremely slowly decaying $4k_\text{F}$ component in the charge-charge correlations, signifying the presence of a one-dimensional Wigner crystal state even for the weakest long-range Coulomb interaction. However, it is not clear how the finite width of a wire and the coupling strength will affect the Wigner crystallization. In fact, one of the purposes of this paper is to find possible answers to this question. At the same time, in the spin correlation function, a finite $2k_\text{F}$ peak was obtained whereas the $4k_\text{F}$ peak was absent.

The experimental observation of a Wigner crystal is a tricky task since it requires extreme conditions such as low temperatures and low densities \cite{Ziani20}. In addition, we show in this paper that it also requires small wire widths. Examining such a fragile crystal requires the use of noninvasive imaging probes. 
In case of 1D infinite systems, true long-range order is forbidden due to thermal and quantum fluctuations. However, in finite systems, physicists have studied 1D Wigner crystals and produced crystalline correlations due to quasi-long-range order. Introducing long-range Coulomb interactions results in a $4k_\text{F}$ oscillation in charge correlations whose decay rate is slower than any power law \cite{Schulz93}. The experimental realization of long-range interactions was made possible in a recent work by Shapir \textit{et al.}\ \cite{Shapir19}, where a few electrons ($<10$) were housed in a carbon nanotube. A second nanotube was used as a scanning probe to measure the spatial charge distribution of electrons in the first nanotube. The measured charge densities were found to exhibit the features of a quasi-Wigner crystal. There is no critical density associated with the onset of the 1D Wigner crystal.

Our aim in this work is to study the effects of confinement and electron correlations on the ground-state properties of a paramagnetic harmonic wire. Unlike our previous work in Ref.\ \onlinecite{Ankush22} we investigate the emergence of quasi-Wigner crystal phases at low density and we examine crossover in the position of the dominant peak in the static structure factor for various wire widths at low density. We employ VMC technique to calculate ground-state properties of the system under consideration. The paper is structured as follows. The model for confinement of electrons in 1D is described in Sec.\ \ref{sec:calculation_details}. The energies for the paramagnetic harmonic wire are presented in Sec.\ \ref{sec:energies}. The dependence of the $2k_\text{F} \rightarrow 4k_\text{F}$ crossover on wire width is discussed in detail in Sec.\ \ref{sec:ssf&pcf}. In Sec.\ \ref{sec:MD}, we present MDs and Tomonaga-Luttinger (TL) parameters $K_{\rho}$ extracted from the MD data around $k \sim k_\text{F}$ as a function of wire width. Our conclusions are drawn in Sec.\ \ref{sec:conclusions}. Hartree atomic units ($\hbar=|e|=m_\text{e}=4\pi\epsilon_0=1$) are used throughout this paper.

\section{Model and calculation details} \label{sec:calculation_details}

The form of Hamiltonian used for simulating a 1D electron fluid is
 \begin{eqnarray}
  \hat{H}=-\frac{1}{2}\sum_{i=1}^{N}\frac{\partial^2}{\partial x_i^2}+\sum_{i<j}\tilde{V}(x_{ij})+\frac{N}{2} V_{\rm Mad},
 \end{eqnarray}
 where $\tilde{V}(x_{ij})$ and $V_{\rm Mad}$ describe the Ewald interaction and Madelung energy, respectively. The Ewald-like interaction for a harmonic wire in a periodic cell of length $L$ is \cite{Saunders94,Lee11}
 \begin{eqnarray}
  \tilde{V}(x_{ij})&=&\sum^{\infty}_{m=-\infty}\bigg[ \frac{\sqrt{\pi}}{2b} e^{{(x_{ij}-mL)}^2/{(4b)}^2} {\rm erfc}\bigg(\frac{|x_{ij}-mL|}{2b}\bigg)\nonumber\\
  & &\hspace{4em} {} -\frac{1}{|x_{ij}-mL|} {\rm erf}\bigg(\frac{|x_{ij}-mL|}{2b} \bigg)\bigg]\nonumber\\
  & &{} +\frac{2}{L}\sum^{\infty}_{n=1} {\rm E}_1[{(bGn)}^2]\cos(Gnx_{ij}),
 \end{eqnarray}
 where $G=2\pi/L$, $b$ is the wire width, and ${\rm E}_1$ is the exponential integral function.

We have considered a 1D HEG in a simulation cell of length $L= 2 N r_\text{s}$ subject to periodic boundary conditions, with $N$ being the total number of electrons in the periodic cell. We keep $N^\uparrow = N^\downarrow$ odd to avoid shell degeneracy effects. A Slater-Jastrow-backflow wave function was used in the VMC calculations \cite{Drummond04, Lopez06}. The orbitals in the Slater determinant were chosen to be plane waves with wave vectors up to $k_\text{F}= \pi/(4r_\text{s})$ for the paramagnetic systems. The VMC method as implemented in the \textsc{casino} code \cite{Needs20} was used to calculate ground-state expectation values. More simulation details can be found in our previous works \cite{Lee11, Vinod18c, Ankush22}.

In a 1D electron gas with an electron-electron interaction potential that is appropriate for a finite wire of width $b$, opposite-spin electrons have a small but nonzero contact probability density. By contrast, in an infinitely thin 1D electron gas, the contact probability density is zero for both same-spin and opposite-spin electrons. QMC calculations for 1D electron gases with finite-width interactions are challenging because the width introduces features on the small length scale $b$ into the wave function near coalescence points. Typically $b$ will be much smaller than other length scales in the problem, such as the density parameter $r_\text{s}$. 
 
One issue that affects diffusion quantum Monte Carlo calculations of finite-width wires is ergodicity; to explore all of configuration space, opposite-spin electrons have to be able to pass one another. However the ``quasinodes'' at opposite-spin coalescence points lead to a very low (but nonzero) probability of electrons passing each other, resulting in very long correlation times. This is much less of an issue in VMC, where there is no need to work in the limit of small time steps, so that electron moves past coalescence points are permitted. A second issue is simply the challenge of creating and optimizing a trial wave function that contains features on the small length scale $b$ as well as the large length scale $L$ of the simulation cell.

\section{Results and discussion}
\label{sec:results}

\subsection{Energies\label{sec:energies}}

\begin{figure}[htbp!]
 \includegraphics[clip,width=.48\textwidth]{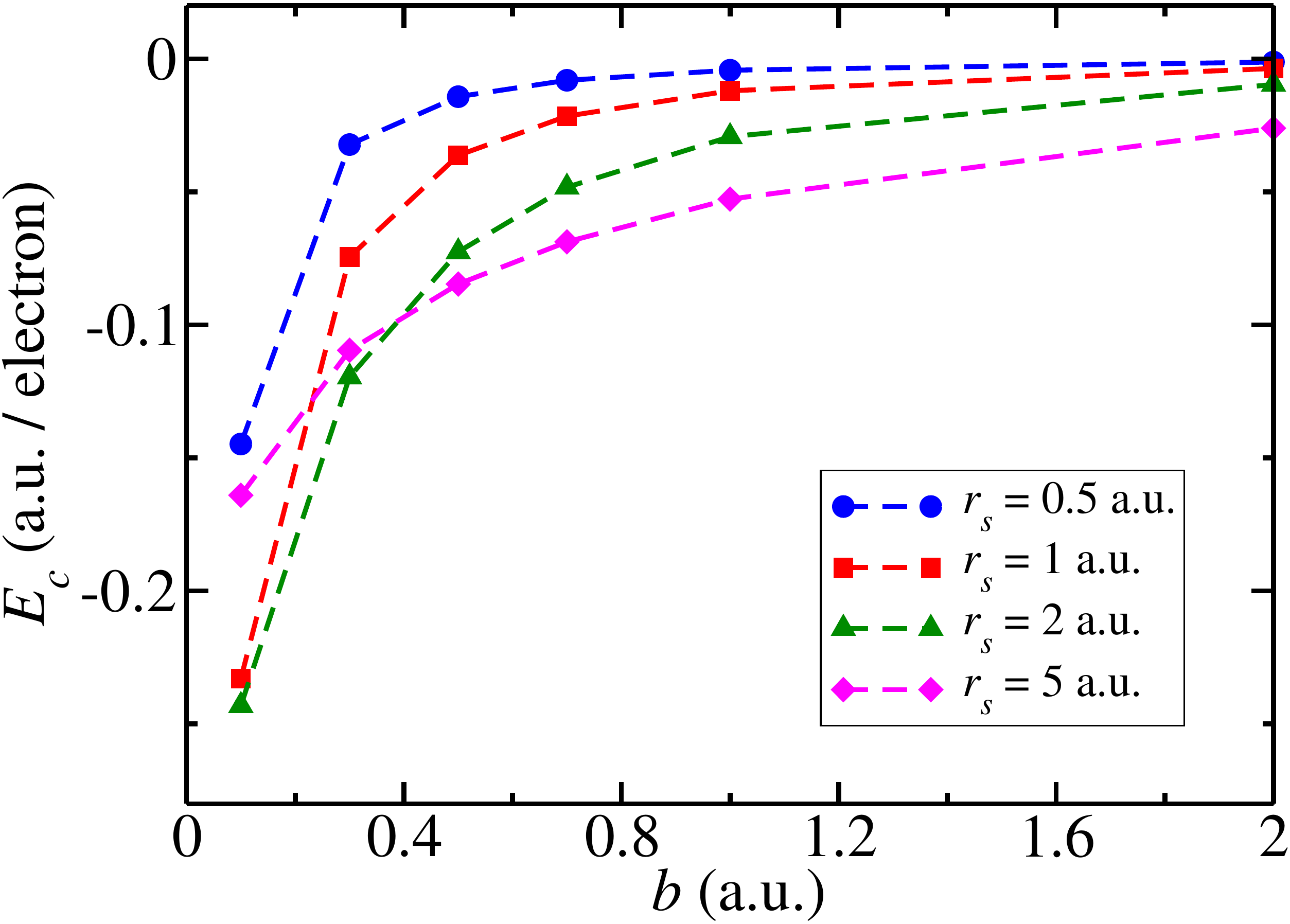}
 \caption{\label{fig:corr_energy} Correlation energy per particle ($E_\text{c}$) as a function of wire width $b$ for $r_\text{s}$ = 0.5, 1, and 2 (top to bottom).}
\end{figure}

The ground-state energy per electron for the harmonic wire is calculated for $r_\text{s}=0.5$, 1, 2, and 5 and for $b=0.1$, 0.3, 0.5, 0.7, 1, 2 a.u. We calculate the energies for $N=30$, 50, 74, 98 electrons and then extrapolating them to obtain the thermodynamic limit. The ground state energy is found to scale with system size as $E(N)=E_{\infty}+ B N^{-2}$ \cite{Lee11} where $B$ and $E_{\infty}$ are fitting parameters. We also calculate the correlation energy per electron using the ground-state energy. These energies are tabulated in Table \ref{tab:table1} of Appendix \ref{apndx1}. The wire width dependence of correlation energy per particle for different values of density parameter $r_\text{s}$ is drawn in Fig.\ \ref{fig:corr_energy}.

\subsection{Structure and correlation function\label{sec:ssf&pcf}}

\begin{figure*}
 \includegraphics[clip,width=.94\textwidth]{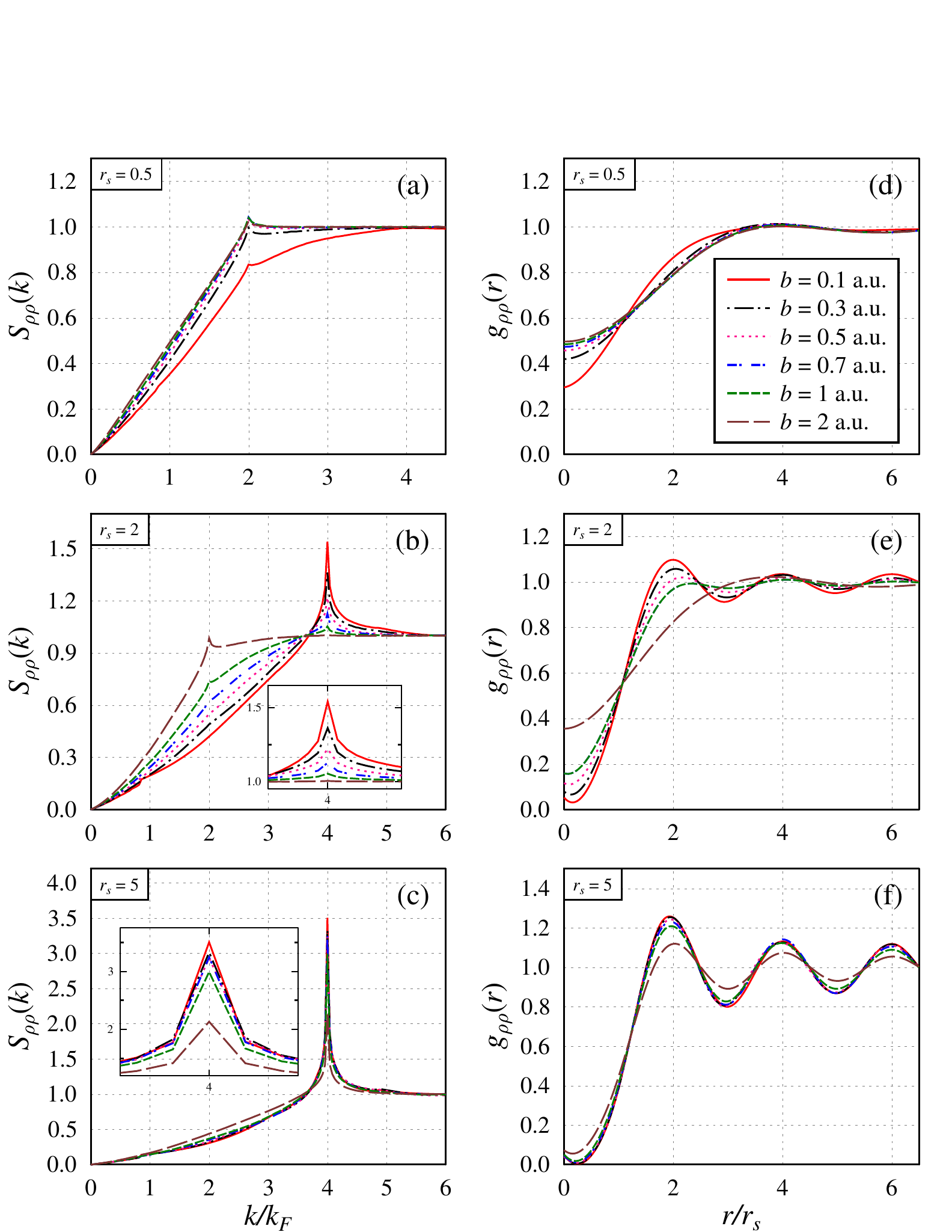}
 \caption{\label{fig:ssfrr_pcfrr_all} Charge structure factor $S_{\rho\rho}(k)$ plotted against $k/k_\text{F}$ (a)--(c) and charge-charge pair correlation function plotted against $r/r_\text{s}$ (d)--(f) for $r_\text{s}=0.5$, 2, and 5 for various $b$ values with $N=98$ electrons. The insets in (b) and (c) show the peak at $k=4k_\text{F}$ in greater detail.}
\end{figure*}

\begin{figure*}
 \includegraphics[clip,width=.94\textwidth]{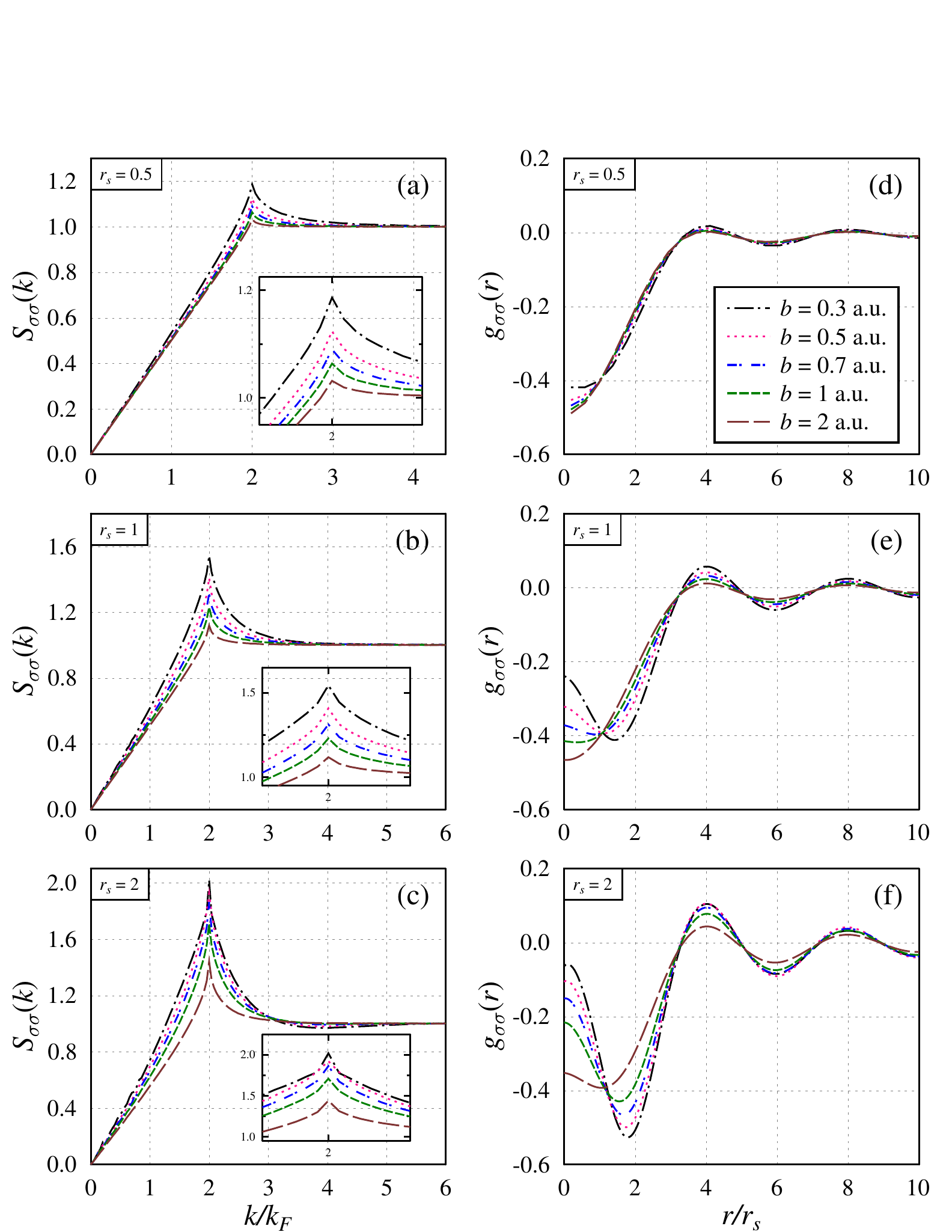}
 \caption{\label{fig:ssfss_pcfss_all} Spin structure factor $S_{\sigma\sigma}(k)$ plotted against $k/k_\text{F}$ (a)--(c) and spin-spin pair correlation function plotted against $r/r_\text{s}$ (d)--(f) for $r_\text{s}=0.5$, 1, and 2 for various $b$ values with $N=98$ electrons. The insets (a), (b), and (c) show the peak at $k=2k_\text{F}$ in greater detail.}
\end{figure*}

\begin{figure}[htbp!]
 \includegraphics[clip,width=.48\textwidth]{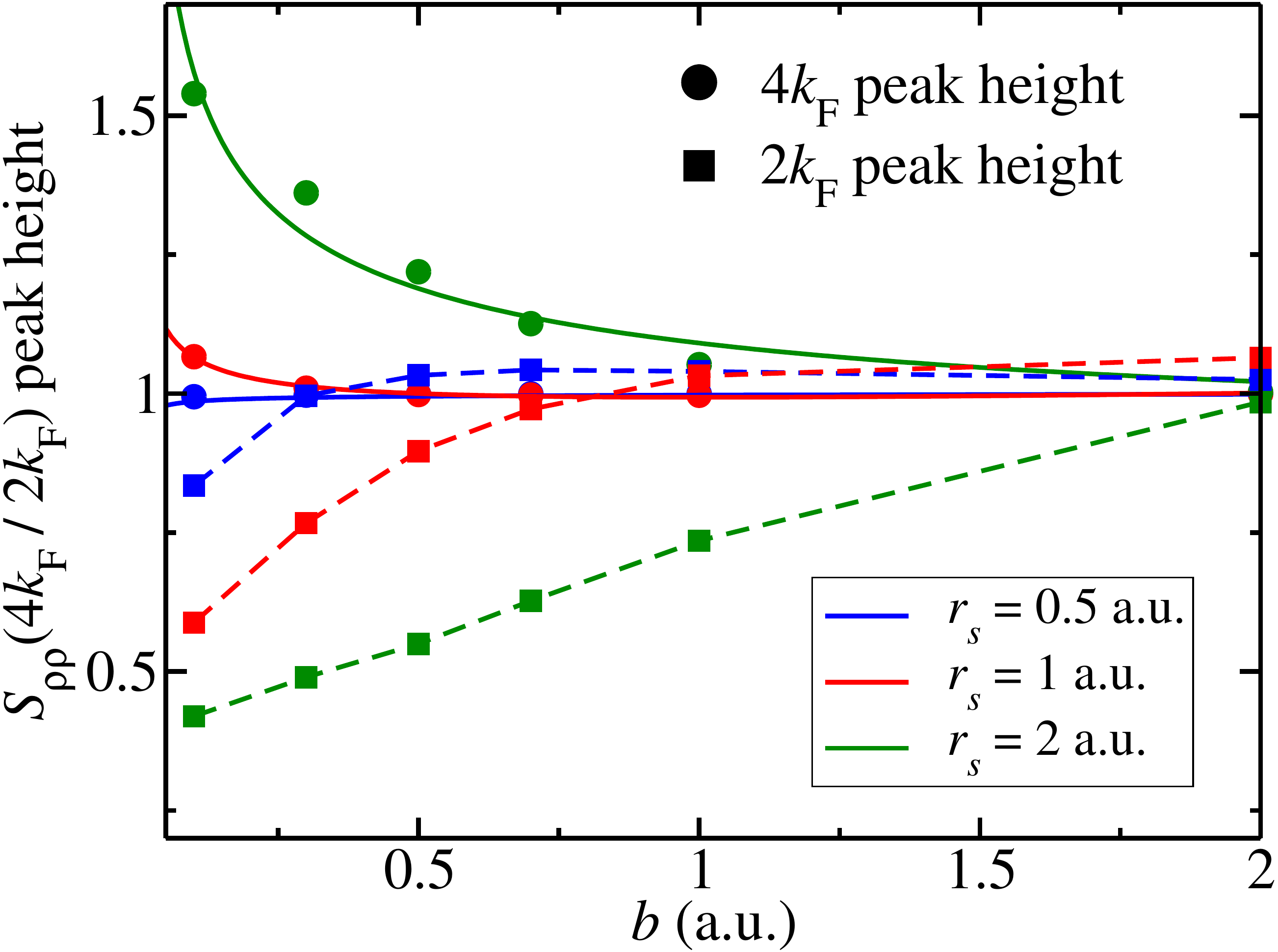} \\
 \includegraphics[clip,width=.48\textwidth]{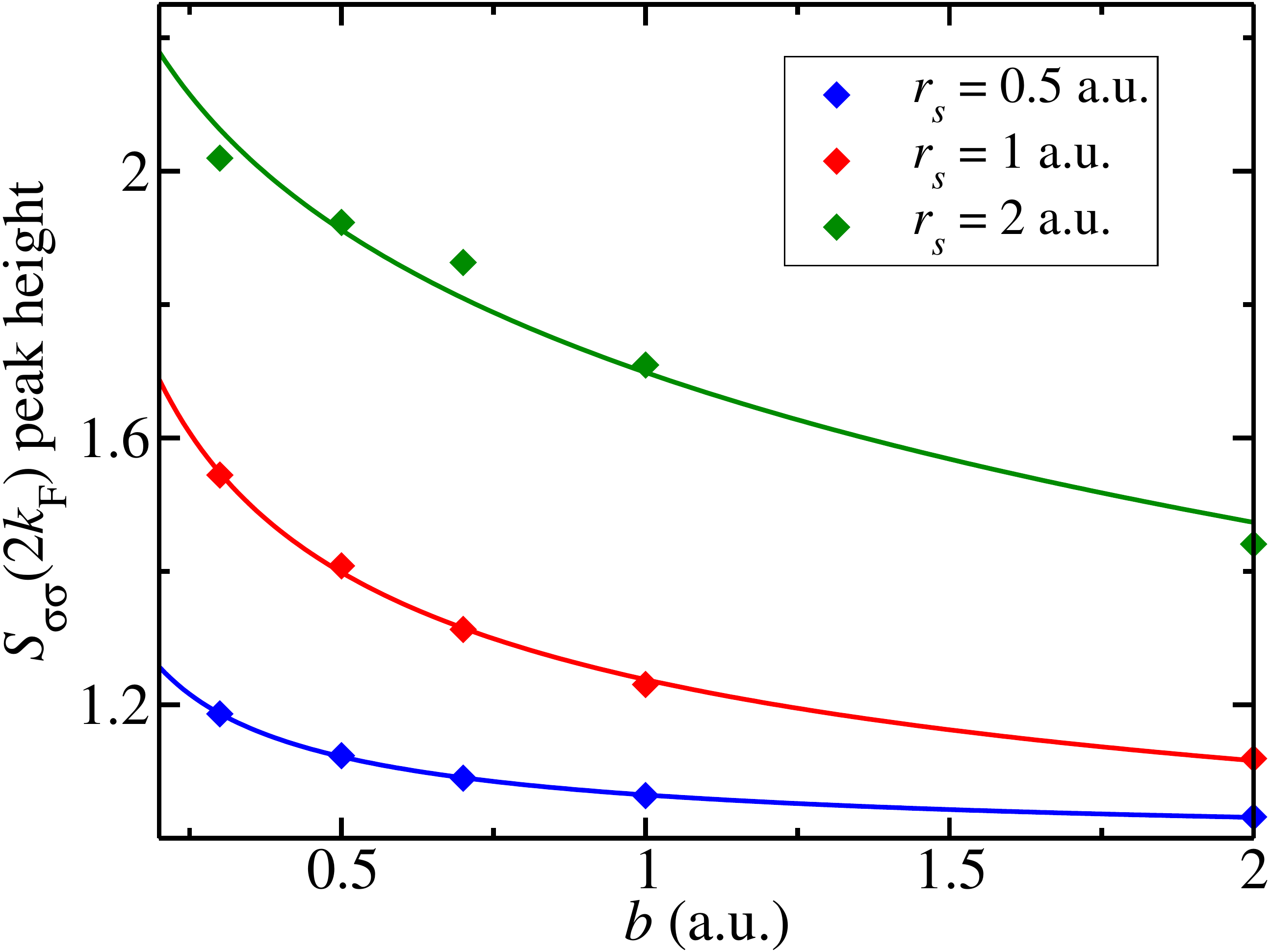}
 \caption{\label{fig:ssf_peak_height_rs_all} (Top) Charge SSF peak height at $k= 4k_\text{F}$ as a function of wire width $b$ for $r_\text{s}=0.5$, 1, and 2 for $N=98$ electrons. The peak heights at $k=4k_\text{F}$ are fitted by Eq.\ (\ref{Eq:cSSF4kf_fit}) and the fits are shown by solid lines, whereas the $2k_\text{F}$ peak heights are joined by dashed lines. (Bottom) Height of spin SSF plotted against $b$ for $r_\text{s}=0.5$, 1, and 2 for $N=98$ electrons and fitted by Eq.\ (\ref{Eq:sSSF2kf_fit}).}
\end{figure}

We calculate the VMC charge and spin static structure factors (SSFs), which are defined as 
\begin{eqnarray} S_{\rho\rho}(k) & = & S_{\uparrow \uparrow}+S_{\uparrow \downarrow} \\
S_{\sigma\sigma}(k) & = & S_{\uparrow \uparrow}-S_{\uparrow \downarrow}, \end{eqnarray}
respectively, for a paramagnetic quantum wire. Here $S_{\uparrow \uparrow}$ and $S_{\uparrow \downarrow}$ are parallel-spin and antiparallel-spin SSFs. The SSF is a crucial property of electron fluids that is used to look into the structural properties of the electron system in reciprocal space.  The charge SSF, also known as the total SSF, describes correlations between electron pairs irrespective of their spins.  The spin PCF describes the difference between pair correlations for parallel- and antiparallel-spin electron pairs.
To study the charge-charge and spin-spin correlation functions in real space, we define alternative quantities, namely charge and spin pair correlation functions (PCFs), as
\begin{eqnarray} g_{\rho \rho} & = & \frac{g_{\uparrow \uparrow}+g_{\uparrow \downarrow}}{2} \\ g_{\sigma \sigma} & = & \frac{g_{\uparrow \uparrow}-g_{\uparrow \downarrow}}{2}, \end{eqnarray} respectively. Here $g_{\uparrow \uparrow}$ and $g_{\uparrow \downarrow}$ are parallel-spin and antiparallel-spin PCFs, respectively. A peak at $k=4k_\text{F}$ in the charge structure factor corresponds to slowly decaying oscillations of period $2 r_\text{s}$, which is the average interparticle spacing. This is a signature of crossover to a ordered phase known as a Wigner crystal. For the case of ferromagnetic infinitely thin wires, the singularity at $k = 4k_\text{F}$ is absent for densities $r_\text{s} < 15$ a.u., but a small feature at $4k_\text{F}$ starts to develop at $r_\text{s} \approx 15$ \cite{Lee11}. The effect of singularity at $4 k_\text{F}$ has been observed recently \cite{Rajesh21} for the case of electron-electron biwires, where the parallel wire effectively provides an extra spin degree of freedom to electrons. In the case of a paramagnetic quasi-quantum wire, it has been verified in previous works \cite{Malatesta00, Casula05} that the dominant $4 k_\text{F}$ peak height grows sublinearly with system size $N$ at low density, which is a signature of a quasi-Wigner crystal. The adjective `quasi' has been used to emphasize the absence of true long-range order due to quantum fluctuations. Further, it has been found that for any fixed wire width, this crossover occurs as one decreases the electron density (i.e., low density with large values of $r_\text{s}$). In this work, we have observed a similar correlation effect by decreasing the wire width at fixed coupling parameter $r_\text{s}$. 

The charge structure factor $S_{\rho\rho}(k)$ is plotted for several
values of $b$ and $r_\text{s}$ for $N=98$ electrons in
Fig.\ \ref{fig:ssfrr_pcfrr_all}. In
Fig.\ \ref{fig:ssfrr_pcfrr_all}(a), $S_{\rho\rho}(k)$ shows a peak at
$k= 2k_\text{F}$ and no peak at $4k_\text{F}$ is present. This leads
to oscillations of period $4 r_\text{s}$ in charge-charge PCF, as can
be seen in Fig.\ \ref{fig:ssfrr_pcfrr_all}(d). It is evident from
Fig.\ \ref{fig:ssfrr_pcfrr_all}(b) that as $b$ decreases for
$r_\text{s}=2$, the $4 k_\text{F}$ peak grows and the $2 k_\text{F}$
peak is suppressed, indicating the crossover ($2 k_\text{F}
\rightarrow 4 k_\text{F}$) to an ordered phase. This crossover leads
to a change in the period of the oscillations in the charge-charge PCF
from $4 r_\text{s}$ to $2 r_\text{s}$, as apparent in
Fig.\ \ref{fig:ssfrr_pcfrr_all}(e). Further, it can be noted from
Fig.\ \ref{fig:ssfrr_pcfrr_all}(c) that the $2 k_\text{F}$ peak is
absent, but decreasing $b$ leads to an increase in the height of the
$4 k_\text{F}$ peak. This is also reflected in a more structured
behavior of $g_{\rho \rho}$, with oscillation period $2 r_\text{s}$ in
Fig.\ \ref{fig:ssfrr_pcfrr_all}(f).

The spin structure factor $S_{\sigma\sigma}(k)$ is plotted for several values of $b$ and $r_\text{s}$ for $N=98$ electrons in Fig.\ \ref{fig:ssfss_pcfss_all}. $S_{\sigma\sigma}(k)$ is found to show a peak at $k=2k_\text{F}$, which grows as $b$ decreases for all the densities considered [see Figs.\ \ref{fig:ssfss_pcfss_all}(a)--(c)]. This is followed by the spin-spin PCF in Figs.\ \ref{fig:ssfss_pcfss_all}(d)--(f), where the amplitude of oscillations of period $4 r_\text{s}$ is enhanced as $b$ decreases. It can also be noted from Figs.\ \ref{fig:ssfss_pcfss_all}(a)--(c) that the peak height increases with increasing $r_\text{s}$ for a given value of $b$. In the present work, we have observed complete decoupling of the charge and spin degrees of freedom in the SSFs. For example, in $r_\text{s} = 2$, there is a crossover from $2 k_\text{F} \rightarrow 4 k_\text{F}$ behavior, with complete disappearance of $2 k_\text{F}$ peak at $b=0.5$, in the charge SSF [Fig.\ \ref{fig:ssfrr_pcfrr_all}(b)], while for the same values of $b$ and $r_\text{s}$ the spin SSF shows a peak at $k=2 k_\text{F}$ [see Fig.\ \ref{fig:ssfss_pcfss_all}(c)]. The spin and charge excitations in the SSFs manifest the complete separation of spin and charge of electrons due to strong Coulomb correlation effects induced by the small transverse confinement width. 

The crossover observed is between (i) a state in which opposite spin electrons can easily slide past each other (large $b/r_\text{s}$; in this case oscillations in both charge and spin PCFs have period $4r_\text{s}$, because correlation is only important for same-spin electrons and negligible for opposite-spin pairs) and (ii) a state in which opposite spin electrons cannot slide past each other (small $b/r_\text{s}$; in this case the system is like a floating antiferromagnetic crystal lattice; the spin PCF has period $4r_\text{s}$, whereas the charge PCF has period $2r_\text{s}$). The ``fluid to crystal'' crossover can be visualized in terms of the behavior of interlocking lattices of spin-up electrons and spin-down electrons, with the crossover occurring at the point at which the lattices start being able to slide past each other.

On the basis of our finite wire width theory \cite{Ankush22}, the SSF is a function of $b/r_\text{s}$. In view of this, and clues from bosonization theory \cite{Schulz93, Fano99}, we propose that the dependence of the peak heights $S_{\rho \rho}(4 k_\text{F},b)$ and $S_{\sigma \sigma}(2 k_\text{F},b)$ on wire width can be represented as 
 \begin{equation}
  \label{Eq:cSSF4kf_fit}
  S_{\rho \rho}(4k_\text{F},b)= a_{0}\frac{L\; r_\text{s}}{b}\exp \left(-4c\sqrt{\ln \frac{L\;r_\text{s}}{b} }\right)+a_{1},
 \end{equation} 
 and 
 \begin{align}
  \label{Eq:sSSF2kf_fit}
  S_{\sigma \sigma}(2k_\text{F},b)=& - a_{2} \left(\sqrt{\ln\frac{L\;r_\text{s}}{b}} + \frac{1}{c} \right) \exp\left(-c \sqrt{\ln \frac{L\;r_\text{s}}{b}}\right) \nonumber\\
  &+a_{3} 
 \end{align}
 where $a_{0}$, $a_{1}$, $a_{2}$, $a_{3}$, and $c$ are fitting parameters and $L=2N r_{\text{s}}$.

Finite wire width effects are studied in Fig.\ \ref{fig:ssf_peak_height_rs_all}. The $4 k_\text{F}$ and $2 k_\text{F}$ peak heights in the charge SSF are plotted against wire width in Fig.\ \ref{fig:ssf_peak_height_rs_all} (top) for $r_\text{s}=0.5$, 1, and 2. The dependence of the peak heights on $b$ is found to be significant for $r_\text{s}=1$ and 2, and for these coupling parameters both the $4 k_\text{F}$ and $2 k_\text{F}$ peaks are present. On the other hand, for $r_\text{s}=0.5$, there is no $4 k_\text{F}$ peak present for any of the wire widths considered and even the $2 k_\text{F}$ peak height depends only weakly on $b$ (at least until $b \ll r_\text{s}$). The $4 k_\text{F}$ peak heights have been fitted to Eq.\ (\ref{Eq:cSSF4kf_fit}), which was deduced on the basis of bosonization and which shows a good representation of the numerically calculated data. The wire width dependence of the $2 k_\text{F}$ peak height in the spin SSF is plotted in Fig.\ \ref{fig:ssf_peak_height_rs_all} (bottom) for $r_\text{s}=0.5$, 1, and 2. It is observed that the peak height in the spin SSF depends on both $r_\text{s}$ and $b$. It is also observed that for a fixed value of electron coupling, the amplitude of the peak height at $k=2 k_\text{F}$ in the spin SSF strongly depends on the confinement width. Further, the $2 k_\text{F}$ peak heights are well fitted by Eq.\ (\ref{Eq:sSSF2kf_fit}). 

It should be noted from Fig.\ \ref{fig:ssf_peak_height_rs_all} that the SSF of the finite-width paramagnetic wire is not expected to converge to that of a ferromagnetic infinitely thin wire in the limit $b \rightarrow 0$, because the latter has only parallel spin correlations whereas the former has both parallel and antiparallel spin correlations. 
Further, the quasi-long-range order in ferromagnetic and paramagnetic electron gases is represented by dominant peaks in the charge structure factor $S_{\rho\rho}(k)$ at $k=2k_\text{F}$ and $k=4k_\text{F}$, respectively, when $b$ is small. The dominant peaks in $S_{\rho\rho}(k)$ for these two cases correspond to the same $2r_\text{s}$ oscillation period in the charge-charge PCF\@. A nonmonotonic shift of the dominant peak from $4k_\text{F}$ to $2k_\text{F}$ for $b\rightarrow 0$ in a paramagnetic wire is not therefore expected. This behavior is consistent with Eq.\ (\ref{Eq:cSSF4kf_fit}), the model formula obtained from the bosonization prediction \cite{Schulz93, Fano99}. It is known that the Lieb-Mattis theorem is not applicable for infinitely thin wires, and in fact the ferromagnetic and paramagnetic states are energetically degenerate in this limit \cite{Lee11}. However, physical observables which do not commute with the Hamiltonian, such as the SSF, need not be same for infinitely thin paramagnetic and ferromagnetic wires. 

\begin{figure}[htbp!]
 \includegraphics[clip,width=.48\textwidth]{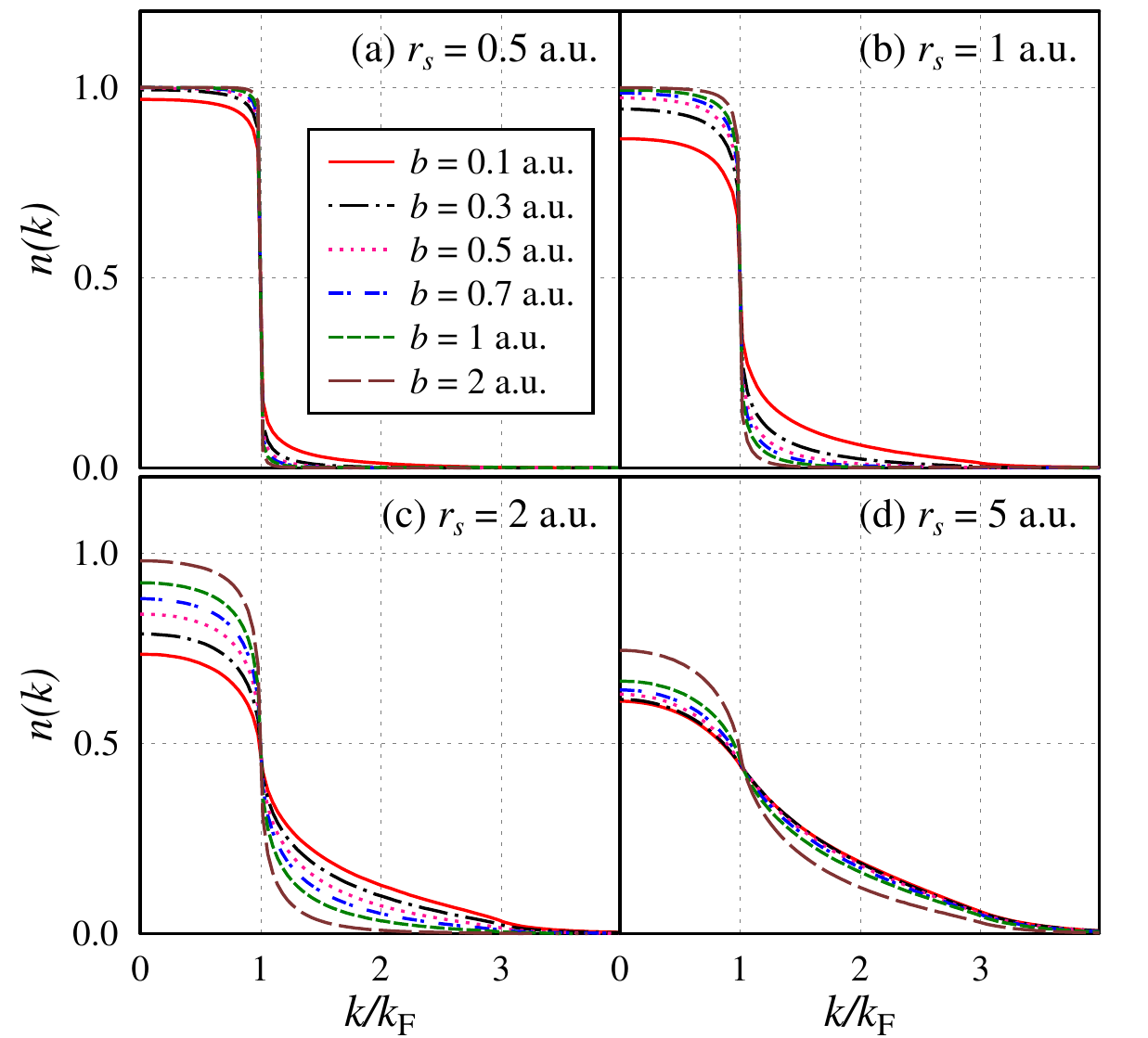}
 \caption{\label{fig:md_rs2} MD against $k/k_\text{F}$ for different values of wire width $b$ and $r_\text{s}$ for $N=98$ electrons.}
\end{figure}

\subsection{Momentum density and TL parameters
 \label{sec:MD}}
 
TL theory suggests that the 1D MD has a peculiar power-law behavior, which is continuous at $k_\text{F}$ with a singular first-order derivative and takes the form $n(k)=n(k_\text{F})+A[{\rm sgn}(k-k_\text{F})]|k-k_\text{F}|^{\alpha}$ close to $k=k_\text{F}$, where $n(k_\text{F})$, $A$, and $\alpha$ are constants \cite{Luttinger_1963, Mattis65}. The method used to estimate the interaction exponent $\alpha$ is discussed in great detail in our previous works \cite{Lee11,Vinod18c,Ankush22}.
 
In Fig.\ \ref{fig:md_rs2}, the MD is plotted for $N=98$ electrons at several values of $b$ and $r_\text{s}$. The effect of wire width on the behavior of the MD is clearly visible in the figure. It can be seen that as $b$ decreases or $r_\text{s}$ increases, the power-law behavior becomes more appreciable. It can also be noted that the value of $n(k=0)$ decreases from 1 and the value of $n(k)$ increases for $k>k_\text{F}$ as $b$ decreases or $r_\text{s}$ increases. 

\begin{figure}[htbp!]
 \includegraphics[clip,width=.48\textwidth]{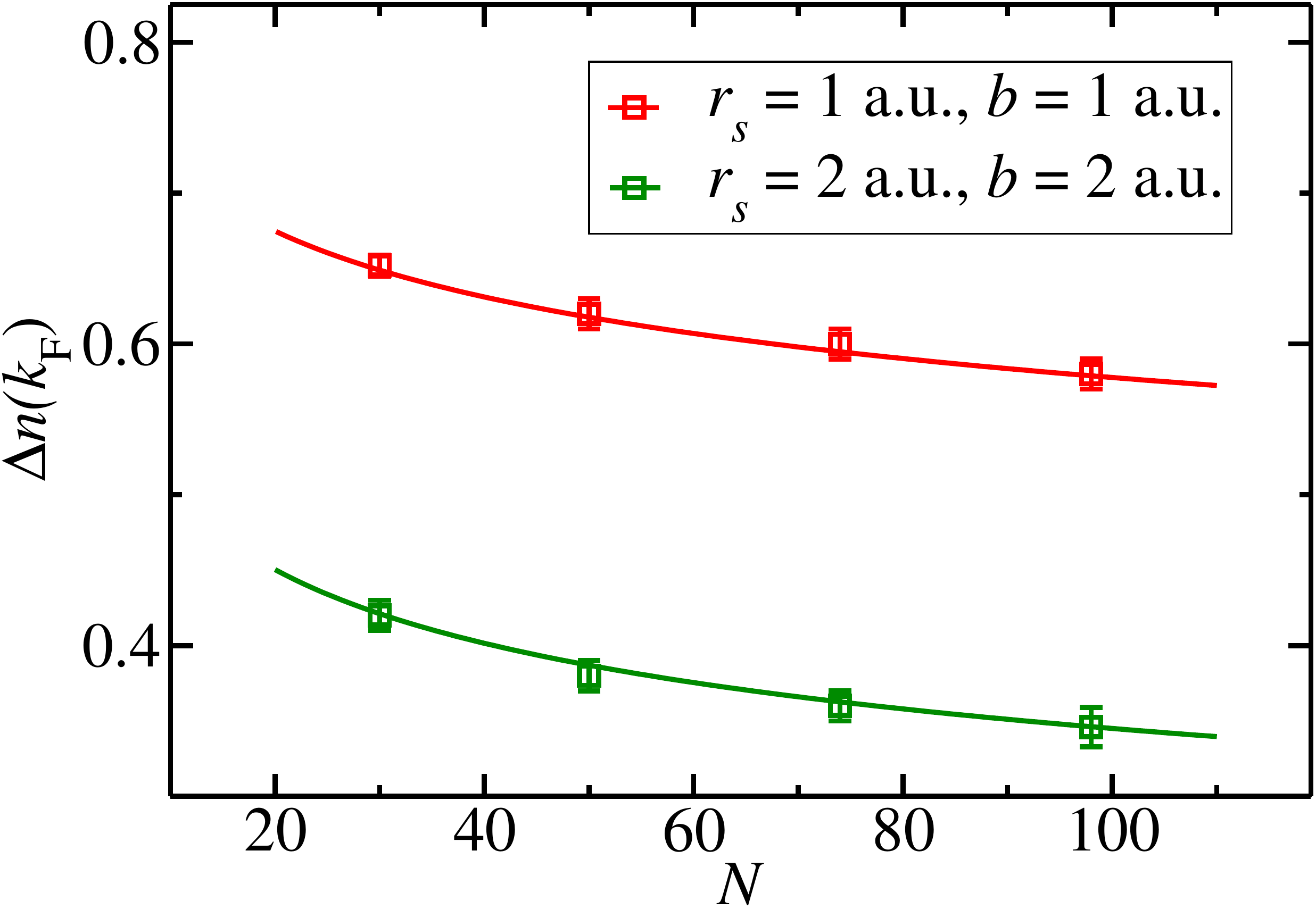}
 \caption{\label{fig:md_jump_with_fit} Jump in the MD at $k=k_\text{F}$ at finite size, plotted against system size and fitted by $\Delta n(k_\text{F})= aL e^{-c {(\ln L)}^{3/2}}$. The fitted parameter values are $a=1.03(2)$ a.u.\ and $c=0.731(2)$ for $r_\text{s}=1$ and $b=1$; and $a=0.93(5)$ a.u.\ and $c=0.777(7)$ for $r_\text{s}=2$ and $b=2$.}
\end{figure}

We also compare our QMC MD data with bosonization results \cite{Schulz93, Fano99}, finding a good fit of the jump in the MD at $k=k_\text{F}$ at finite size by the asymptotic scaling function predicted by bosonization theory:
\begin{equation}
    \Delta n(k_\text{F}) \sim L e^{-c {(\ln L)}^{3/2}}. \label{eq:bos_MD}
\end{equation}
Figure \ref{fig:md_jump_with_fit} shows that Eq.\ (\ref{eq:bos_MD}) provides quite a good numerical representation of the system-size scaling of the jump in the MD at $k=k_\text{F}$.

\begin{figure}[htbp!]
 \includegraphics[clip,width=.48\textwidth]{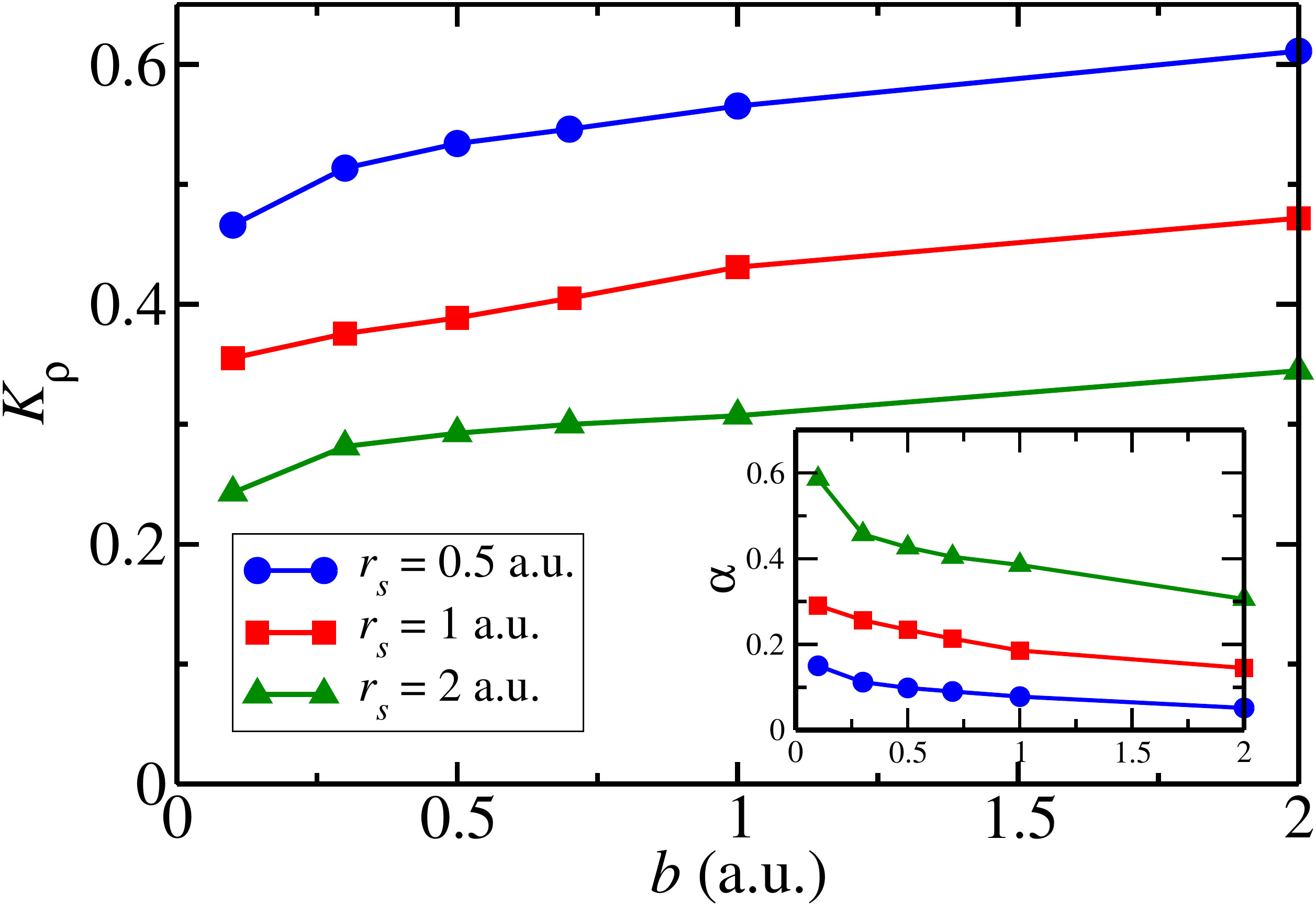}
 \caption{\label{fig:tl_par_all} TL parameter $K_\rho$ against wire width $b$ for different values of the density parameter $r_\text{s}$ as indicated in the plot. The TL parameter is obtained by fitting MD data to appropriate fitting functions to obtain the exponent $\alpha$ and $K_\rho$. The exponent $\alpha$ is plotted against $b$ in the inset. }
\end{figure}

The TL parameter $K_\rho$ for repulsive interactions ($K_\rho<1$) and exponent $\alpha$ are plotted against $b$ for $r_\text{s}=0.5$, 1, and 2 in Fig.\ \ref{fig:tl_par_all}. It has been observed that as $b$ decreases or $r_\text{s}$ increases, $K_\rho$ decreases and the interacting electrons become more strongly correlated. Hence it is found that $K_\rho$ significantly depends on the confinement width $b$ and $r_\text{s}$. 

\section{Conclusions}\label{sec:conclusions}

We have studied the wire width dependence of the ground-state properties of quasi-1D quantum wires for several electron densities. We analyze the effects of adjusting the wire width on the charge structure factor, where a crossover from a peak at $2k_\text{F}$ to a peak at $4k_\text{F}$ was observed as $b$ decreases. The crossover indicates the onset of a quasi-Wigner crystalline state in the 1D electron fluid. The PCF is also found to show a more structured behavior as $b$ decreases, which again suggests a crossover to a quasi-Wigner crystal phase. We find the crossover does not appear for all values of $r_\text{s}$ and $b$, a result at variance with the bosonization findings of Schulz in Ref.\ \cite{Schulz93}. We also fit the dominant peaks of the charge and spin structure factors assuming fitting functions on the basis of our finite wire width theory and clues from bosonization, giving a good fit of the VMC data. The crossover occurs only for those density and wire width values which bring strong correlations in the system.

The correlation energy as a function of $b$ and $r_\text{s}$ is represented by a formula which fits our VMC data perfectly.
The MD $n(k)$ is found to show a power law behavior. We find that $n(k)$ at $k=0$ reduces from $1$, whereas $n(k)$ increases beyond $k_\text{F}$ as $b$ decreases or $r_\text{s}$ increases. We find the TL liquid exponent $\alpha$ by fitting the MD data to an appropriate fitting function near $k_\text{F}$. The TL parameter $K_{\rho}$ is found to depend significantly on the wire width at a constant $r_\text{s}$.

\begin{acknowledgments}
One of the authors (A.G.)\ would like to thank MHRD for financial support. K.N.P.\ acknowledges the National Academy of Sciences of India, Prayagraj for partial financial assistance. R.S.\ acknowledges SERB for providing NPDF grant PDF/2021/000546. V.A.\ acknowledges financial support from DST-SERB Grant No.\ EEQ/2019/000528, and the National PARAM Supercomputing Facility at C-DAC Pune for computational resources.
\end{acknowledgments}

\onecolumngrid

\appendix   
\section{}\label{apndx1}
The thermodynamic limit of the VMC ground state energy per electron $E_{\infty}$ and the correlation energy per electron $E_\text{c}$ are calculated by extrapolation from calculations performed at several different values of system size $N$ at different values of $b$ and $r_\text{s}$; these are tabulated in Table \ref{tab:table1}.

\squeezetable
\begin{table*}[htbp!]
  \begin{center}
 \caption{\label{tab:table1} VMC energies in a.u.\ per electron [$E(N)$] for $N = 30$, 50, 74, and 98 for a paramagnetic harmonic wire. $E_{\infty}$ and $E_\text{c}$ denote the ground-state energy per electron extrapolated to the thermodynamic limit and the correlation energy per electron respectively.}
 \begin{ruledtabular}
  \begin{tabular}{ccccccc}
   ($r_\text{s}, b$) & $E(30)$ & $E(50)$& $E(74)$& $E(98)$& $E_{\infty}$ & $E_\text{c}$\\ \hline
   (0.5, 0.1)& $-$0.937394(8) & $-$0.935599(7) & $-$0.935042(5) & $-$0.934791(5) & $-$0.93455(2) & $-$0.14477(2)  \\
   (0.5, 0.3)& $-$0.336230(3) & $-$0.334478(2) & $-$0.333931(2) & $-$0.333725(2) & $-$0.333473(7)  & $-$0.032178(7) \\
   (0.5, 0.5)& $-$0.127807(2) & $-$0.126094(1) & $-$0.1255490(9) & $-$0.1253469(9) & $-$0.125100(9) & $-$0.014176(9) \\
   (0.5, 0.7)& $-$0.0145572(9) & $-$0.0128579(7) & $-$0.0123206(6) & $-$0.0121214(5)& $-$0.011877(8) &  $-$0.008008(8) \\   
   (0.5, 1.0)& 0.0853158(5) & 0.0869805(4) & 0.0875067(3) & 0.0877014(3) & 0.087942(8) & $-$0.004262(8)  \\
   (0.5, 2.0)& 0.225663(2) & 0.2272236(1) & 0.22772541(9) & 0.22791182(8) & 0.228151(5) & $-$0.001176(5) \\   
   (1.0, 0.1)& $-$0.899453(5) & $-$0.898910(4) & $-$0.898695(3) & $-$0.898505(3) & $-$0.89847(6) & $-$0.23295(6) \\
   (1.0, 0.3)& $-$0.478440(2) & $-$0.477901(2) & $-$0.477725(2) & $-$0.477671(1) & $-$0.477589(4) & $-$0.074459(4) \\
   (1.0, 0.5)& $-$0.328023(1) & $-$0.327485(1) & $-$0.327323(1) & $-$0.3272540(9) & $-$0.327179(4) & $-$0.036280(4) \\
   (1.0, 0.7)& $-$0.245460(1) & $-$0.2449252(8) & $-$0.2447628(7) & $-$0.2446985(6) & $-$0.244622(2) & $-$0.021498(2) \\
   (1.0, 1.0)& $-$0.1710782(7) & $-$0.1705563(5) & $-$0.1703899(4) & $-$0.1703292(4) & $-$0.170254(3) & $-$0.011984(3) \\     
   (1.0, 2.0)& $-$0.0610819(3) & $-$0.0605826(2) & $-$0.0604240(2) & $-$0.0603641(1) & $-$0.060292(3) & $-$0.003585(3) \\   
   (2.0, 0.1)& $-$0.687882(6) & $-$0.687592(7) & $-$0.687429(6) & $-$0.687309(2) & $-$0.68726(3) & $-$0.24309(3) \\      
   (2.0, 0.3)& $-$0.428893(1) & $-$0.428709(1) & $-$0.4286485(9)& $-$0.4285848(9) & $-$0.42858(2) & $-$0.11954(2) \\ 
   (2.0, 0.5)& $-$0.321088(1) & $-$0.3209185(8) & $-$0.3208670(7) & $-$0.3208426(6) & $-$0.320819(2) & $-$0.072522(2) \\   
   (2.0, 0.7)& $-$0.2583233(8) & $-$0.2581581(7) & $-$0.2581048(6) & $-$0.2580852(5) & $-$0.258061(1) & $-$0.048212(1) \\ 
   (2.0, 1.0)& $-$0.2006410(7) & $-$0.2004732(5) & $-$0.2004218(5) & $-$0.2004024(4) & $-$0.2003781(3) & $-$0.0292266(3) \\
   (2.0, 2.0)& $-$0.1146782(3) & $-$0.1145145(2) & $-$0.1144626(2) & $-$0.1144422(2) & $-$0.114419(1) & $-$0.009582(1) \\ 
   (5.0, 0.1)& $-$0.393883(2) & $-$0.393808(3) & $-$0.393718(2) & $-$0.3935256(8)& $-$0.39352(6) & $-$0.16401(6) \\   
   (5.0, 0.3)& $-$0.284488(3) & $-$0.283437(9) & $-$0.284425(1)  & $-$0.2843393(4)& $-$0.28433(5) & $-$0.10955(5) \\
   (5.0, 0.5)& $-$0.234260(4) & $-$0.234304(6) & $-$0.234296(8) & $-$0.2342142(4)& $-$0.23420(3) & $-$0.08464(3) \\   
   (5.0, 0.7)& $-$0.201875(3) & $-$0.2019512(7) & $-$0.2019229(7) & $-$0.2019149(2)& $-$0.201911(9) & $-$0.068787(9)  \\     
   (5.0, 1.0)& $-$0.168710(3) & $-$0.1687108(6) & $-$0.1686853(6) & $-$0.1686792(2)& $-$0.168670(4) & $-$0.052681(4) \\  
   (5.0, 2.0)& $-$0.1102775(8) & $-$0.1102383(7) & $-$0.1102273(5) & $-$0.1102212(1)& $-$0.1102150(4) & $-$0.0260165(4) 
  \end{tabular}
  \end{ruledtabular}
  \end{center}
\end{table*}%

\clearpage
\twocolumngrid



\begin{thebibliography}{0}%
\makeatletter
\providecommand \@ifxundefined [1]{%
 \@ifx{#1\undefined}
}%
\providecommand \@ifnum [1]{%
 \ifnum #1\expandafter \@firstoftwo
 \else \expandafter \@secondoftwo
 \fi
}%
\providecommand \@ifx [1]{%
 \ifx #1\expandafter \@firstoftwo
 \else \expandafter \@secondoftwo
 \fi
}%
\providecommand \natexlab [1]{#1}%
\providecommand \enquote  [1]{``#1''}%
\providecommand \bibnamefont  [1]{#1}%
\providecommand \bibfnamefont [1]{#1}%
\providecommand \citenamefont [1]{#1}%
\providecommand \href@noop [0]{\@secondoftwo}%
\providecommand \href [0]{\begingroup \@sanitize@url \@href}%
\providecommand \@href[1]{\@@startlink{#1}\@@href}%
\providecommand \@@href[1]{\endgroup#1\@@endlink}%
\providecommand \@sanitize@url [0]{\catcode `\\12\catcode `\$12\catcode
  `\&12\catcode `\#12\catcode `\^12\catcode `\_12\catcode `\%12\relax}%
\providecommand \@@startlink[1]{}%
\providecommand \@@endlink[0]{}%
\providecommand \url  [0]{\begingroup\@sanitize@url \@url }%
\providecommand \@url [1]{\endgroup\@href {#1}{\urlprefix }}%
\providecommand \urlprefix  [0]{URL }%
\providecommand \Eprint [0]{\href }%
\providecommand \doibase [0]{https://doi.org/}%
\providecommand \selectlanguage [0]{\@gobble}%
\providecommand \bibinfo  [0]{\@secondoftwo}%
\providecommand \bibfield  [0]{\@secondoftwo}%
\providecommand \translation [1]{[#1]}%
\providecommand \BibitemOpen [0]{}%
\providecommand \bibitemStop [0]{}%
\providecommand \bibitemNoStop [0]{.\EOS\space}%
\providecommand \EOS [0]{\spacefactor3000\relax}%
\providecommand \BibitemShut  [1]{\csname bibitem#1\endcsname}%
\let\auto@bib@innerbib\@empty
\end{thebibliography}%


\begin{thebibliography}{36}
 \bibitem{Giuliani_2005} G.\ F.\ Giuliani and G.\ Vignale, \textit{Quantum Theory of the Electron Liquid} (Cambridge University Press, Cambridge, 2005). 
 \bibitem{Giamarchi04} T.\ Giamarchi, \textit{Quantum Physics in One Dimension} (Clarendon, Oxford, 2004). 
 \bibitem{Auslaender05} O.\ M.\ Auslaender, H.\ Steinberg, A.\ Yakoby, Y.\ Tserkovnyak, B.\ I.\ Halperin, K.\ W.\ Baldwin, L.\ N.\ Pfeiffer, and K.\ W.\ West, Science {\bf 308}, 88 (2005). 
 \bibitem{Bockrath99} M.\ Bockrath, D.\ H.\ Cobden, J.\ Lu, A.\ G.\ Rinzler, R.\ E.\ Smalley, L.\ Balents, and P.\ L.\ McEuen, Nature {\bf 397}, 598 (1999). 
 \bibitem{Steinberg08} H.\ Steinberg, G.\ Barak, A.\ Yacoby, L.\ N.\ Pfeiffer, K.\ W.\ West, B.\ I.\ Halperin, and K.\ Le Hur, Nat.\ Phys.\ {\bf 4}, 116 (2008). 
 \bibitem{Deshpande08} V.\ V.\ Deshpande and M.\ Bockrath, Nat.\ Phys.\ {\bf 4}, 314 (2008).
 \bibitem{Pecker13}S.\ Pecker, F.\ Kuemmeth, A.\ Secchi, M.\ Rontani, D.\ C.\ Ralph, P.\ L.\ McEuen, and S.\ Ilani, Nat.\ Phys.\ \textbf{9}, 576 (2013).
 \bibitem{Shapir19}I.\ Shapir, A.\ Hamo, S.\ Pecker, C.\ Moca, \'O.\ Legeza, G.\ Zarand, and S.\ Ilani, Science \textbf{364}, 870 (2019). 
 \bibitem{Ziani20}  N.\ T.\ Ziani, F.\ Cavaliere, K.\ G.\ Becerra, and M.\ Sassetti, Crystals, \textbf{11}, 20 (2020). 
 \bibitem{Tomonaga_1950} S.\ Tomonaga, Prog.\ Theor.\ Phys.\ {\bf 5}, 544 (1950).
 \bibitem{Luttinger_1963} J.\ M.\ Luttinger, J.\ Math.\ Phys.\ {\bf 4}, 1154 (1963).
 \bibitem{Haldane81} F.\ D.\ M.\ Haldane, Phys.\ Rev.\ Lett.\ {\bf 47}, 1840 (1981).
 \bibitem{Schulz83}H.\ J.\ Schulz, J.\ Phys.\ C \textbf{16}, 6769 (1983). 
 \bibitem{Gold90} A.\ Gold and A.\ Ghazali, Phys.\ Rev.\ B \textbf{41}, 7626 (1990).
 \bibitem{Hu93}B.\ Y.-K.\ Hu and S.\ Das Sarma, Phys.\ Rev.\ B, \textbf{48}, 5469 (1993).
 \bibitem{Friesen80} W.\ I.\ Friesen and B.\ Bergersen, J.\ Phys.\ C {\bf 13}, 6627 (1980). 
 \bibitem{Hu90}G.\ Y.\ Hu and R.\ F.\ O'Connell, Phys.\ Rev.\ B, \textbf{42}, 1290 (1990).
 \bibitem{Sun93}Y.\ Sun and G.\ Kirczenow, Phys.\ Rev.\ B, \textbf{47}, 4413 (1993). 
 \bibitem{Poilblanc97} D.\ Poilblanc, S.\ Yunoki, S.\ Maekawa, and E.\ Dagotto, Phys.\ Rev.\ B {\bf 56}, R1645 (1997).
 \bibitem{Renu14} R.\ Bala, R.\ K.\ Moudgil, S.\ Srivastava, and K.\ N.\ Pathak, Eur.\ Phys.\ J.\ B \textbf{87}, 5 (2014).
 \bibitem{Vinod18} V.\ Ashokan, R.\ Bala, K.\ Morawetz, and K.\ N.\ Pathak, Eur.\ Phys.\ J.\ {\bf 91}, 29 (2018).
 \bibitem{Morawetz18} K.\ Morawetz, V.\ Ashokan, R.\ Bala, and K.\ N.\ Pathak, Phys.\ Rev.\ B \textbf{97}, 155147 (2018).
 \bibitem{Vinod20} V.\ Ashokan, R.\ Bala, K.\ Morawetz, and K.\ N.\ Pathak, Phys.\ Rev.\ B \textbf{101}, 075130 (2020) 
 \bibitem{Camels97}L.\ Camels and A.\ Gold, Europhys.\ Lett.\ \textbf{39}, 539 (1997).
 \bibitem{Fano99} G.\ Fano, F.\ Ortolani, A.\ Parola, and L.\ Ziosi, Phys.\ Rev.\ B {\bf 60}, 15654 (1999).
 \bibitem{Li19} Z.-H.\ Li, J.\ Phys.\: Condens.\ Matter \textbf{31}, 255601 (2019).
 \bibitem{Malatesta00} A.\ Malatesta and G.\ Senatore, J.\ Phys.\ IV {\bf 10}, 5 (2000).
 \bibitem{Casula05}M.\ Casula and G.\ Senatore, ChemPhysChem \textbf{6}, 1902 (2005).  
 \bibitem{Lee11} R.\ M.\ Lee and N.\ D.\ Drummond, Phys.\ Rev.\ B {\bf 83}, 245114 (2011).
 \bibitem{Vinod18c}V.\ Ashokan, N.\ D.\ Drummond, and K.\ N.\ Pathak, Phys.\ Rev.\ B {\bf 98}, 125139 (2018).
 \bibitem{Ankush22} A.\ Girdhar, V.\ Ashokan, N.\ D.\ Drummond, K.\ Morawetz, and K.\ N.\ Pathak, Phys.\ Rev.\ B \textbf{105}, 115140 (2022).
 \bibitem{Casula06}M.\ Casula, S.\ Sorella, and G.\ Senatore, Phys.\ Rev.\ B \textbf{74}, 245427 (2006).
 \bibitem{Shulenburger08} L.\ Shulenburger, M.\ Casula, G.\ Senatore, and R.\ M.\ Martin, Phys.\ Rev.\ B {\bf 78}, 165303 (2008).
 \bibitem{Schulz93}H.\ J.\ Schulz, Phys.\ Rev.\ Lett.\ \textbf{71}, 1864 (1993).
 \bibitem{Saunders94} V.\ R.\ Saunders, C.\ Freyria-Fava, R.\ Dovesi, and C.\ Roetti, Comput.\ Phys.\ Commun.\ \textbf{84}, 156 (1994).
 \bibitem{Drummond04} N.\ D.\ Drummond, M.\ D.\ Towler, and R.\ J.\ Needs, Phys.\ Rev.\ B {\bf 70}, 235119 (2004). 
 \bibitem{Lopez06} P.\ L{\'o}pez R\'{\i}os, A.\ Ma, N.\ D.\ Drummond, M.\ D.\ Towler, and R.\ J.\ Needs, Phys.\ Rev.\ E {\bf 74}, 066701 (2006). 
 \bibitem{Needs20} R.\ J.\ Needs, M.\ D.\ Towler, N.\ D.\ Drummond, P.\ L\'opez R\'{\i}os, and J.\ R.\ Trail, J.\ Chem.\ Phys.\ \textbf{152}, 154106 (2020). 
 \bibitem{Rajesh21} R.\ O.\ Sharma, N.\ D.\ Drummond, V.\ Ashokan, K.\ N.\ Pathak, and K.\ Morawetz,  Phys.\ Rev.\ B {\bf 104}, 035149 (2021). 
\bibitem{Mattis65} D.\ C.\ Mattis and E.\ H.\ Lieb, J.\ Math.\ Phys.\ {\bf 6}, 304 (1965). 
\end{thebibliography}
\end{document}